\documentclass[12pt,a4paper]{article}
\usepackage{amsfonts,latexsym}
\usepackage{amsmath,amssymb}
\usepackage{graphicx,color}

\renewcommand{\thefootnote}{\fnsymbol{footnote}}

\begin{document}

\vspace{12mm}

\begin{center}
{{{\Large {\bf New massive conformal gravity }}}}\\[10mm]

{Yun Soo Myung\footnote{e-mail address: ysmyung@inje.ac.kr}}\\[8mm]

{Institute of Basic Sciences and Department  of Computer Simulation, Inje University Gimhae 621-749, Korea\\[0pt]}

\end{center}
\vspace{2mm}

\begin{abstract}
We investigate the new massive conformal gravity which is not
invariant under conformal transformations, in comparison  to the
massive conformal gravity. We find five polarization modes of
gravitational waves propagating on the Minkowski spacetimes.  The
stability of Minkowski spacetimes is guaranteed if the mass squared
is not negative and the linearized Ricci tenor is employed to
describe a massive spin-2 graviton. However, the small Schwarzschild
black hole is unstable against the $s$-mode massive graviton
perturbations.
\end{abstract}
\vspace{5mm}



\vspace{1.5cm}

\hspace{11.5cm}{Typeset Using \LaTeX}
\newpage
\renewcommand{\thefootnote}{\arabic{footnote}}
\setcounter{footnote}{0}


\section{Introduction}
Recently, massive conformal gravity (MCG) was proposed as another
massive gravity model~\cite{Faria:2013hxa,Myung:2014aia}. This model
is composed of a conformally coupled scalar to Einstein-Hilbert term
and Weyl-squared term ($C_{\mu\nu\rho\sigma}C^{\mu\nu\rho\sigma}$)
which are invariant under conformal transformations. The other
aspects of the action including the MCG have  been studied
extensively since eighteen years
ago~\cite{Flanagan:1996gw,Mannheim:2001kk,Mannheim:2005bfa,Flanagan:2006ra,Bouchami:2007en,Maldacena:2011mk}.

It was argued that the MCG (\ref{MCG}) is a promising quantum
gravity model with eight degrees of freedom (DOF) because the
conformal symmetry restricts the number of counter-terms arising
from the perturbative quantization of the metric
tensor~\cite{tHooft:2011aa}. On the contrary, Stelle has shown  that
the combination of $ R+\alpha
C_{\mu\nu\rho\sigma}C^{\mu\nu\rho\sigma}+\beta R^2$ is necessary to
improve the perturbative properties of Einstein
gravity~\cite{Stelle:1976gc}. This describes also 8(=2+5+1) DOF
without scalar. If $\alpha \beta \not=0$, the renormalizability was
achieved but the unitarity was violated, showing that the
renormalizability is not compatible with the  unitarity. Although
the Weyl-squared term of providing the massive spin-2 graviton
improves the ultraviolet divergence, it induces ghost excitations
which spoil the unitarity simultaneously. In this approach,  the
price one has to pay for making the theory renormalizable is the
loss of unitarity.   If one excludes the Ricci-squared term, there
is no massive spin-0 corrections. Thus, the MCG including the
Weyl-squared term solely might not be a candidate for a proper
quantum gravity model when one urges to use the metric formalism. Up
to now, there is no obvious way to enhance the renormalizability
without violating the unitarity in fourth-order gravity.

On the other hand, all undesirable issues of Fierz-Pauli massive
gravity with 5 DOF appear when one takes the massless limit of
$m^2\to 0$ to match with a massless spin-2 graviton with 2
DOF~\cite{Hinterbichler:2011tt}. Surely, there is a mismatch in DOF:
3 versus 2, which was known to be the van Dam-Vletman-Zakharov
(vDVZ) discontinuity~\cite{vanDam:1970vg,zak}. Even though massive
gravity has survived as the de Rham-Gebadadze-Tolley (dRGT)
gravity~\cite{deRham:2010ik,deRham:2010kj}, the dRGT gravity has
still the problems of superluminal propagation and local acausality
which mean that it could not be a UV-complete fundamental theory of
gravity~\cite{Deser:2012qx,Deser:2013qza}. Hence, in order to have a
better situation,  one may propose  a new direction of  separating
the Einstein gravity from the massive gravity: ``The Einstein
gravity is described by the metric perturbation $h_{\mu\nu}$ (metric
formalism), whereas the massive gravity is described by the
linearized Ricci tensor $\delta R_{\mu\nu}$ (Ricci tensor
formalism)." In this view, one does not need to recover the Einstein
gravity by taking the massless limit of the massive gravity and,
thus there is no the vDVZ discontinuity.

In this sense, it is worth noting that the MCG might be   a
candidate for massive gravity model with 6(=5+1) DOF if one
expresses  a massive spin-2 graviton in terms of the linearized
Ricci tensor $\delta R_{\mu\nu}$ instead of the metric perturbation
$h_{\mu\nu}$. Here, we are free from linear and non-linear
(Boulware-Deser) ghosts because the linearized Einstein equation
becomes a second-order differential equation and there is no the
vDVZ discontinuity. More importantly, it is well known that the
Riemann tensor $R_{\mu\rho\nu\sigma}$ which causes relative
acceleration between test particles is the only measurable
field~\cite{Alves:2009eg}. One uses  a null-tetrad basis to compute
the Newman-Penrose quantities~\cite{Newman:1961qr} in terms of the
irreducible parts of $R_{\mu\rho\nu\sigma}$ (the Weyl tensor
$C_{\mu\rho\nu\sigma}$, the traceless Ricci tensor
$\tilde{R}_{\mu\nu}$,  and the Ricci scalar $R$). According to the
analysis in~\cite{Eardley:1974nw}, there are six polarization modes
of gravitational waves (GWs) in the most general case which will be
detected  by feasible experiments~\cite{Aasi:2013wya}. For the new
massive gravity in three dimensions, one has  two
polarizations~\cite{Moon:2011gg}. Also, it would be interesting  to
see Ref.~\cite{Fleury:2013uqa} for cosmologically different
observations of the Weyl and Ricci tensors: Ricci-dominated lensing
includes  large beams of CMB measurements and Weyl-dominated lensing
detects narrow beams of  SN observations.

The author has shown that the MCG might not be a promising model of
massive gravity~\cite{Myung:2014aia}. The reason is that the
non-propagation of the linearized Ricci scalar ($\delta R=0$) is a
strong condition to achieve a massive gravity theory at the
linearized level when one uses the Ricci tensor formalism.  However,
one could not obtain $\delta R=0$ because of the conformal symmetry.
Adding the Einstein-Hilbert term $R$ breaks conformal symmetry in
the MCG, leading to the new massive conformal gravity (NMCG).

In this work, we wish to test  the NMCG as a candidate of massive
gravity model with 6(=5+1) DOF. For this purpose, we find five
polarization modes of gravitational waves propagating on the
Minkowski spacetimes in addition to conformal scalar. We investigate
the stability of Minkowski spacetimes as well as the Schwarzschild
black hole by  using the NMCG. The Minkowski spacetimes is stable
against the conformal scalar and linearized Ricci tensor
perturbation if the mass squared is not negative ($m^2\ge 0$). On
the other hand, the small Schwarzschild black hole is unstable
against the $s$-mode massive graviton perturbations under the
condition of $0<m/\sqrt{2}\le 1/r_0$ with $r_0$ the black hole
horizon size.

\section{New massive conformal gravity}
We start with  the new massive conformal gravity (NMCG) action
\begin{eqnarray}S_{\rm NMCG}=\frac{1}{32 \pi }\int d^4 x\sqrt{-g}
\Big[R-\alpha\Big(\phi^2R+
6\partial_\mu\phi\partial^\mu\phi\Big)-\frac{1}{2m^2}C^{\mu\nu\rho\sigma}C_{\mu\nu\rho\sigma}\Big].
\label{NMCG}
\end{eqnarray}
Without the  Einstein-Hilbert term, (\ref{NMCG}) reduces to the
massive conformal gravity (MCG)
\begin{eqnarray}S_{\rm MCG}=\frac{1}{32 \pi }\int d^4 x\sqrt{-g}
\Big[\alpha\Big(\phi^2R+
6\partial_\mu\phi\partial^\mu\phi\Big)-\frac{1}{m^2}C^{\mu\nu\rho\sigma}C_{\mu\nu\rho\sigma}\Big],
\label{MCG}
\end{eqnarray}
which  is invariant under the full conformal transformations
as~\cite{Faria:2013hxa}
\begin{equation} \label{cft}
g_{\mu\nu} \to \Omega^2(x)g_{\mu\nu},~~\phi \to \frac{\phi}{\Omega}.
\end{equation}
Here $\Omega(x)$ is an arbitrary function of the spacetime
coordinates.  Since its conformal transformed action of the
Einstein-Weyl gravity has been ruled out by Solar System
observations (the deflection of light) as a fourth-order gravity in
addition to ghost state problem~\cite{Flanagan:2006ra}. This implies
that  it is meaningless to study the Newtonian approximation to the
MCG. Hence, one would be better to consider the MCG as a massive
gravity model. In this case, one has a difficulty to obtain the
massive graviton equation due to the conformal symmetry, if one does
not require  a relation of $\varphi=\delta
R/6m^2$~\cite{Myung:2014aia}.

Adding the Einstein-Hilbert term $R$ breaks conformal symmetry in
the MCG (\ref{MCG}), leading to the NMCG (\ref{NMCG}). Hence, it
seems that the idea of imposing exact conformal symmetry as a
criterion to choose possible actions of massive gravity does not
work one would expects~\cite{Faria:2013hxa}. Furthermore, there is
no  way to avoid ghost states if one considers any fourth-order
gravity models. Both (\ref{NMCG}) and (\ref{MCG}) become unhealthy
gravity theories when one uses the metric perturbation. However, the
theory becomes a healthy massive gravity theory if one expresses  a
massive graviton in terms of the linearized Ricci tensor $\delta
R_{\mu\nu}$ instead of metric perturbation $h_{\mu\nu}$. This is so
because the linearized Einstein equation becomes a second-order
differential equation.  Similarly, the linearized topologically
massive gravity became a first-order theory if one introduces a
linearized Einstein tensor $\delta G_{\mu\nu}$~\cite{Li:2008dq}. In
this case, one does not need to introduce a chiral (critical)
gravity to avoid the ghost states when one uses the linearized
Einstein tensor. In the case of new massive
gravity~\cite{Bergshoeff:2009hq}, the linearized filed equation
around the Minkowski vacuum is given by $(\square-m^2)\delta
R_{\mu\nu}=0$ which is considered as a boosted-up version of the
Fierz-Pauli equation
[$(\square-m^2)h_{\mu\nu}=0$]~\cite{Bergshoeff:2013vra}.

At this stage, we note that even though the last of Weyl-squared
term  is invariant under conformal transformations (\ref{cft}), we
include it as a second-order term because this term provides a
unique way of achieving a massive gravity model without a massive
spin-0 graviton.

The Einstein equation is derived from (\ref{NMCG}) as
\begin{equation} \label{nequa1}
G_{\mu\nu}=\alpha
\Big[\phi^2G_{\mu\nu}+g_{\mu\nu}\nabla^2(\phi^2)-\nabla_\mu\nabla_\nu(\phi^2)+6\partial_\mu\phi\partial_\nu\phi-3(\partial\phi)^2g_{\mu\nu}\Big]+\frac{1}{m^2}B_{\mu\nu},
\end{equation}
where the Einstein tensor  is given by \begin{equation}
G_{\mu\nu}=R_{\mu\nu}-\frac{1}{2} Rg_{\mu\nu}
\end{equation}
and the Bach tensor $B_{\mu\nu}$  takes the form
\begin{eqnarray} \label{equa2}
B_{\mu\nu}&=& 2 \Big(R_{\mu\rho\nu\sigma}R^{\rho\sigma}-\frac{1}{4}
R^{\rho\sigma}R_{\rho\sigma}g_{\mu\nu}\Big)-\frac{2}{3}
R\Big(R_{\mu\nu}-\frac{1}{4} Rg_{\mu\nu}\Big) \nonumber \\
&+&
\nabla^2R_{\mu\nu}-\frac{1}{6}\nabla^2Rg_{\mu\nu}-\frac{1}{3}\nabla_\mu\nabla_\nu
R.
\end{eqnarray}
Its trace is zero  ($B^\mu~_\mu=0$). On the other hand, the scalar
equation is given by
\begin{equation} \label{ascalar-eq}
\nabla^2\phi-\frac{1}{6}R\phi=0.
\end{equation}
Taking the trace of (\ref{nequa1}) leads to
\begin{equation} \label{ricciz}
R=0
\end{equation}
which might be used to simplify the scalar equation
(\ref{ascalar-eq}) as a massless scalar equation
\begin{equation} \label{tscalar-eq}
\nabla^2\phi=0.
\end{equation}

\section{Polarization modes of GWs  in Minkowski
spacetimes} By setting
\begin{equation}
\bar{R}_{\mu\nu\rho\sigma}=0,~~\bar{R}_{\mu\nu}=0,~~\bar{R}=0,~~\bar{\phi}=\sqrt{\frac{1}{2\alpha}},
\end{equation}
 we have the Minkowski  background as
 \begin{equation}
 \bar{g}_{\mu\nu}=\eta_{\mu\nu}={\rm diag}(-+++).
 \end{equation}
 In order to develop  polarization modes of gravitational waves (GWs), we
consider the Ricci tensor and Ricci scalar as first-order functions
of the metric perturbation $h_{\mu\nu}$ in
\begin{equation}
g_{\mu\nu}=\eta_{\mu\nu}+h_{\mu\nu}. \end{equation} Then, we have
perturbed Ricci tensor, Ricci scalar, and conformal scalar around
the background quantities
\begin{equation}
R_{\mu\nu}=0+\delta R_{\mu\nu},~~R=0+\delta
R,~~\phi=\bar{\phi}(1+\varphi).
\end{equation}
We immediately obtain the non-propagation of the Ricci scalar from
(\ref{ricciz}) as
\begin{eqnarray} \label{riscalar-eq}
\delta R=0.
\end{eqnarray}
Considering (\ref{riscalar-eq}), the perturbed equations are derived
from (\ref{tscalar-eq}) and (\ref{nequa1}) as
\begin{eqnarray}
\label{lis-eq1}&&\square \varphi=0, \\
\label{lis-eq2}&&\square \delta R_{\mu\nu}-\frac{m^2}{2}\Big(\delta
R_{\mu\nu}-2\partial_\mu\partial_\nu \varphi\Big)=0.
\end{eqnarray}
A plane wave solution to (\ref{lis-eq1}) is given by
\begin{equation}
\varphi=\varphi_0 e^{iq_\mu x^\mu},~~q_\mu q^\mu=0.
\end{equation}
On the other hand, a plane solution to (\ref{lis-eq2}) takes the
form~\cite{Alves:2009eg} \begin{equation} \label{plane-sol} \delta
R_{\mu\nu}=A_{\mu\nu} e^{(iqz-\omega t)}+B_{\mu\nu}e^{i(kz-\omega
t)}+{\rm c.c.},
\end{equation}
where \begin{equation} \label{wavevec} A_{\mu\nu}=-2\varphi_0 q_\mu
q_\nu,~~q=\omega,~~k=\sqrt{\omega^2-\frac{m^2}{2}} \end{equation}
with the frame choice of $q_\mu=(\omega,0,0,q)$ and
$k_\mu=(\omega,0,0,k)$ to describe GWs propagating in the $+z$
direction. So, all the quantities are functions of $t$ and $z$ only
in this section.

We find all explicit forms of $\delta R_{\mu\nu}$ as
\begin{eqnarray}
&&\delta R_{tt}=-2\varphi_0 \omega^2 e^{(iqz-\omega
t)}+B_{tt}e^{i(kz-\omega t)}+{\rm c.c.},\\
&&\delta R_{tz}=-2\varphi_0 \omega q e^{(iqz-\omega
t)}+B_{tz}e^{i(kz-\omega t)}+{\rm c.c.},\\
&&\delta R_{zz}=-2\varphi_0 q^2 e^{(iqz-\omega
t)}+B_{zz}e^{i(kz-\omega t)}+{\rm c.c.}
\end{eqnarray}
and all other components satisfy
\begin{equation}
\delta R_{ij}=B_{ij}e^{i(kz-\omega t)}+{\rm c.c.},
\end{equation}
where $i,j=x,y,z$. Since there is no further constraints on
$B_{\mu\nu}$, all components of the Riemann tensor in the tetrad
basis are given by \begin{equation}
R_{lklk}=0,~~R_{lml\tilde{m}}\not=0,~~R_{lklm}\not=0,~~R_{lkl\tilde{m}}\not=0,
\end{equation}
where the first term comes from the non-propagation condition of
$\delta R=0$ (\ref{riscalar-eq}).
 They  correspond to
the Newman-Penrose quantities of Riemann-tensor
as~\cite{Eardley:1974nw}
\begin{equation}
\Psi_2=0,~~\Psi_3\not=0,~~\Psi_4\not=0,~~\Phi_{22}\not=0
\end{equation}
whose helicity values are assigned to be $s=\{0,\pm1,\pm2,0\}$,
respectively when considering Lorentz rotations.  These all describe
a massive spin-2 graviton with 5 DOF propagating on  the Minkowski
spacetimes.   Together with a conformal scalar $\varphi$, the NMCG
could describe 6 DOF. If one uses the metric formalism with
$h_{\mu\nu}$, then the NMCG describes 8 DOF (5 for massive graviton,
1 for conformal scalar, and 2 for massless
graviton)~\cite{Stelle:1976gc}. This is another difference in DOF
between two formalisms, in addition to the essential difference
between second-order and fourth-order linearized equations.

Let us compare the NMCG with the MCG.  In the case of the MCG, we
had a relation of $\delta R=6m^2 \varphi$ which implies that
$(\square-m^2)\{\varphi,\delta R\}=0$. In this case, one has $
\delta R\not=0\to \Psi_2\not=0$. Hence, $\Psi_2$ is regarded as a
key observational gravitational wave to discriminate between NMCG
and MCG. In addition, one has  two polarizations of $\Phi_{12}$ and
$\Phi_{22}$ for the new massive gravity in three
dimensions~\cite{Moon:2011gg}.

Finally, we would like to mention the stability of the Minkowski
spacetimes. As was shown in (\ref{wavevec}), we have the dispersion
relation for the linearized Ricci tensor propagation
\begin{equation}
\omega^2=k^2+\frac{m^2}{2}
\end{equation}
which implies that $\omega$ is always real for $m^2\ge 0$, leading
to the stability of the Minkowski spacetimes against  a massive
plane wave (\ref{plane-sol}). However, for the tachyonic mass of
$m^2=-M^2$, one may have  other dispersion relation
\begin{equation}
\omega^2=k^2-\frac{M^2}{2},
\end{equation}
which implies the existence of a characteristic wave vector
$k_*=M/\sqrt{2}$ making $\omega=0$.  Hence, for $k>k_*$, the wave
dominates and thus, one achieves the stability (the frequency
$\omega$ is still real and one has oscillations) even though the
tachyon mass appears. However, for $k<k_*$, the tachyonic mass
dominates and thus, the frequency $\omega$ becomes purely imaginary
($\omega=-i\Omega$). This  implies an exponentially growing mode of
$e^{\Omega t}$ which shows an unstable mode. This is an origin of
the tachyonic instability in the Minkowski spacetimes which is
similar to the Jeans' instability in Newtonian
gravity~\cite{Harmark:2007md}.  The Jeans instability is also
another pendant for  the five-dimensional black string
instability~\cite{Gregory:1993vy}. In the next section, however, we
will show an instability of $s$-mode of the linearized Ricci tensor
propagating around the Schwarzschild black hole spacetimes even for
$m^2>0$.

\section{Instability of massive spin-2 graviton \\ around the Schwarzschild black hole}

Considering the background ansatz
\begin{equation}
\bar{R}_{\mu\rho\nu\sigma}\not=0,~~\bar{R}_{\mu\nu}=0,~~\bar{R}=0,~~\bar{\phi}=\sqrt{\frac{1}{2\alpha}},
\end{equation}
Eq. (\ref{nequa1}) and (\ref{tscalar-eq}) provide   the
Schwarzschild black hole solution \begin{equation} \label{schw}
ds^2_{\rm Sch}=\bar{g}_{\mu\nu}dx^\mu
dx^\nu=-f(r)dt^2+\frac{dr^2}{f(r)}+r^2d\Omega^2_2
\end{equation}
with the metric function \begin{equation} \label{num}
f(r)=1-\frac{r_0}{r}.
\end{equation}
It is easy to show that the Schwarzschild  black hole (\ref{schw})
is also the solution to the Einstein equation of $G_{\mu\nu}=0$ in
Einstein gravity. The event horizon appears at $r=r_0$.

We  introduce the metric and scalar perturbations around the
Schwarzschild  black hole
\begin{eqnarray} \label{m-p}
g_{\mu\nu}=\bar{g}_{\mu\nu}+h_{\mu\nu},~~\phi=\bar{\phi}(1+\varphi)=\sqrt{\frac{1}{2\alpha}}(1+\varphi).
\end{eqnarray}
The linearized Einstein equation around the Schwarzschild black hole
is given by
\begin{eqnarray} \label{nlin-eq}
&&m^2\Big[\frac{1}{2}\delta G
_{\mu\nu}+\bar{g}_{\mu\nu}\bar{\nabla}^2\varphi-\bar{\nabla}_\mu\bar{\nabla}_\nu\varphi\Big]
\\ \nonumber
&& =\Big[\bar{\nabla}^2\delta
G_{\mu\nu}+2\bar{R}_{\rho\mu\sigma\nu}\delta G^{\rho\sigma}\Big]
-\frac{1}{3}\Big[\bar{\nabla}_\mu\bar{\nabla}_\nu-\bar{g}_{\mu\nu}\bar{\nabla}^2
\Big] \delta R,
\end{eqnarray}
where the linearized Einstein tensor, Ricci tensor, and Ricci scalar
are expressed in terms of $h_{\mu\nu}$
\begin{eqnarray}
\delta G_{\mu\nu}&=&\delta R_{\mu\nu}-\frac{1}{2} \delta
R\bar{g}_{\mu\nu},
\label{ein-t} \\
\delta
R_{\mu\nu}&=&\frac{1}{2}\Big(\bar{\nabla}^{\rho}\bar{\nabla}_{\mu}h_{\nu\rho}+
\bar{\nabla}^{\rho}\bar{\nabla}_{\nu}h_{\mu\rho}-\bar{\nabla}^2h_{\mu\nu}-\bar{\nabla}_{\mu}
\bar{\nabla}_{\nu}h\Big), \label{ricc-t} \\
\delta R&=& \bar{g}^{\mu\nu}\delta R_{\mu\nu}= \bar{\nabla}^\mu
\bar{\nabla}^\nu h_{\mu\nu}-\bar{\nabla}^2 h \label{Ricc-s}
\end{eqnarray}
with $h=h^\rho~_\rho$.

Considering (\ref{tscalar-eq}),  its linearized scalar equation is
still given by
\begin{equation}\label{nlsca}
\bar{\nabla}^2\varphi=0
\end{equation}
whose scalar has a propagating wave being free from unstable
modes~\cite{Myung:2014nua}. Taking the trace of the linearized
Einstein equation and using (\ref{nlsca}), one has
\begin{equation}
-\frac{m^2}{2}\delta R=0
\end{equation}
which implies the non-propagation of linearized Ricci scalar
\begin{equation}
\delta R=0
\end{equation}
for  $m^2\not=0$. We note that $\delta R=0$ is confirmed by
linearizing $R=0$ (\ref{ricciz}) directly. The choice of $\delta
R=0$ reflects why we prefer the MCG (\ref{MCG}) to the NMCG
(\ref{NMCG}) as a starting action in this work.  We stress again
that if one does not break conformal symmetry, one could not achieve
the non-propagation of the linearized Ricci scalar.   Plugging
$\delta R=0$ and (\ref{nlsca}) into Eq. (\ref{nlin-eq}) leads to the
linearized Einstein equation for the linearized Ricci tensor
\begin{equation} \label{slin-eq}
\bar{\nabla}^2\delta R_{\mu\nu}+ 2\bar{R}_{\rho\mu\sigma\nu}\delta
R^{\rho\sigma}(=-\Delta_L\delta R_{\mu\nu})=\frac{m^2}{2}\Big[\delta
R_{\mu\nu}-2\bar{\nabla}_\mu\bar{\nabla}_\nu\varphi\Big],
\end{equation}
where $\Delta_L$ is the Lichnerowicz operator. Eq. (\ref{slin-eq})
is still difficult to be solved because of coupling $\delta
R_{\mu\nu}$ and $\varphi$.  In the Minkowski background, Eq.
(\ref{slin-eq}) reduces to (\ref{lis-eq2}).

 Fortunately, Eq.
(\ref{slin-eq}) could be expressed compactly by introducing
$\delta\tilde{ R}_{\mu\nu}=\delta
R_{\mu\nu}-2\bar{\nabla}_\mu\bar{\nabla}_\nu \varphi$ as
\begin{equation} \label{tildeR-eq} \bar{\nabla}^2\delta
\tilde{R}_{\mu\nu}+2\bar{R}_{\rho\mu\sigma\nu}\delta
\tilde{R}^{\rho\sigma}=\frac{m^2}{2}\delta \tilde{R}_{\mu\nu},
\end{equation}
where  we used an important relation~\cite{Myung:2014nua}
\begin{eqnarray} \label{lich-ricci}
\Delta_L \delta \tilde{R}_{\mu\nu}=\Delta_L \delta R_{\mu\nu}.
\end{eqnarray}
In proving (\ref{lich-ricci}), we have used the  relation
\begin{eqnarray}
\Delta_L (\bar{\nabla}_\mu\bar{\nabla}_\nu \varphi)
=-\frac{1}{2}\Big(\bar{\nabla}_\mu\bar{\nabla}_\nu+\bar{\nabla}_\mu\bar{\nabla}_\nu\Big)\bar{\nabla}^2\varphi=0,
 \label{relation}
\end{eqnarray}
 where in the second line, we used the linearized scalar equation (\ref{nlsca}).
It is important to note that Eq. (\ref{tildeR-eq}) could describe
the massive spin-2 field (5 DOF) propagating around the
Schwarzschild black hole, because $\delta\tilde{R}_{\mu\nu}$
satisfies the transverse and traceless  condition
\begin{eqnarray}
\bar\nabla^{\mu}\delta\tilde{R}_{\mu\nu}~ =\delta\tilde{R}~=~0,
\end{eqnarray}
where the contracted Bianchi identity  was used to prove the
transverse condition. After replacing
\begin{eqnarray}
\delta\tilde{R}_{\mu\nu}\to\delta R_{\mu\nu},~~\frac{m^2}{2}\to m^2,
\end{eqnarray}
we find the linearized Ricci tensor equation~\cite{Myung:2013doa}
\begin{equation} \label{ricciten-eq} \bar{\nabla}^2\delta
R_{\mu\nu}+2\bar{R}_{\rho\mu\sigma\nu}\delta
R^{\rho\sigma}=m^2\delta R_{\mu\nu},
\end{equation}
where one has found  unstable modes of $e^{\Omega t}$ for
\begin{equation} \label{unst-con1} 0<m<\frac{{\cal
O}(1)}{r_0}
\end{equation}
in fourth-order gravity. The Schwarzschild black hole found from the
dRGT gravity is also unstable against the $s$-mode of metric
perturbations $h_{\mu\nu}$~\cite{Babichev:2013una,Brito:2013wya}.

Similarly, we find unstable modes for
\begin{equation} \label{unst-con2} 0<\frac{m}{\sqrt{2}}<\frac{{\cal
O}(1)}{r_0}
\end{equation}
in the NMCG.

In order to find the origin of this instability,  we consider a
five-dimensional black string described by~\cite{Gregory:1993vy}
\begin{equation}
ds^2_{\rm BS}=ds^2_{\rm Sch}+dz^2,
\end{equation}
we have perturbation along an extra direction of the $z$-axis
\begin{eqnarray}
h_{AB}=e^{ikz}e^{\tilde{\Omega} t}\left(
\begin{array}{cc}
h_{\mu\nu} & 0 \cr 0 & 0
\end{array}
\right). \label{evenpt}
\end{eqnarray}
Using the transverse-traceless gauge condition of $\bar{\nabla}^\mu
h_{\mu\nu}=0$ and $h=0$, the linearized equation to the Einstein
equation of $R_{AB}=0$ reduces to
\begin{equation} \label{hmn-eq}
\bar{\nabla}^2h_{\mu\nu}+2\bar{R}_{\rho\mu\sigma\nu}h^{\rho\sigma}=k^2h_{\mu\nu},
\end{equation}
which describes a massive spin-2 graviton with 5 DOF propagating
around the Schwarzschild black hole. One has found a long wavelength
perturbation of $0<k<k_c\sim \frac{1}{r_0}$ along  $z$-axis, which
gives us an unstable mode of $e^{\tilde{\Omega}t}$. This is the
Gregory-Laframme instability in the black string theory. Comparing
(\ref{ricciten-eq}) with (\ref{hmn-eq}), one finds that they are the
same by replacing $\delta R_{\mu\nu}$ and $m^2$ by $h_{\mu\nu}$ and
$k^2$.  This implies that the instability of the black hole in the
NMCG arises from the massiveness of $m^2\not=0$ where the geometry
of extra $z$ dimension trades for mass~\cite{Deser:2013qza}.

\section{Discussions}
We have studied the new massive conformal gravity as a candidate for
massive gravity model with 6 DOF. Using the Ricci tensor formalism,
we have found five polarization modes of gravitational waves
propagating on the Minkowski spacetimes, in addition to a single
conformal scalar.

The stability of Minkowski spacetimes is guaranteed if the mass
squared is not negative ($m^2\ge 0$) and the linearized Ricci tenor
was employed to describe a massive spin-2 graviton.  However, the
small Schwarzschild black hole is unstable against the $s$-mode
massive graviton perturbations for $0<m/\sqrt{2}<1/r_0$ which
corresponds to the positive mass-squared case of $m^2>0$. This
instability is a common feature of the Schwarzschild black hole
found from massive gravity
theories~\cite{Babichev:2013una,Brito:2013wya}. Comparing it with
the five-dimensional black string instability, the instability of
the black hole in the NMCG arises from the massiveness of
$m^2\not=0$ where the geometry of extra $z$ dimension trades for
mass.

Consequently, the new massive conformal gravity is regarded as a
promising massive gravity model with 6 DOF if one uses the Ricci
tensor formalism instead of the metric formalism.  The main
difference between new massive conformal gravity (\ref{NMCG}) and
massive conformal gravity (\ref{MCG}) is that the former has no
$\Psi_2$ gravitational wave due to the non-propagation of Ricci
scalar, while the latter has $\Psi_2$ gravitational wave when one
requires a relation between conformal scalar and Ricci scalar
($\varphi=\delta R/6m^2$) additionally. However, the numbers of DOF
are the same six for two massive gravity theories.

 \vspace{2cm}

{\bf Acknowledgments}
 \vspace{1cm}

The author thanks Taeyoon Moon for helpful  discussions.  This work
was supported by the National Research Foundation of Korea (NRF)
grant funded by the Korea government (MEST)
(No.2012-R1A1A2A10040499).


\begin{thebibliography}{99}
\bibitem{Faria:2013hxa}
  F.~F.~Faria,
  arXiv:1312.5553 [gr-qc].

\bibitem{Myung:2014aia}
  Y.~S.~Myung,
  Phys.\ Lett.\ B {\bf 730}, 130 (2014)  [arXiv:1401.1890 [gr-qc]].

\bibitem{Flanagan:1996gw}
  E.~E.~Flanagan and R.~M.~Wald,
  Phys.\ Rev.\ D {\bf 54}, 6233 (1996)  [gr-qc/9602052].

\bibitem{Mannheim:2001kk}
  P.~D.~Mannheim,
  Int.\ J.\ Mod.\ Phys.\ D {\bf 12}, 893 (2003)  [astro-ph/0104022].


\bibitem{Mannheim:2005bfa}
  P.~D.~Mannheim,
   Prog.\ Part.\ Nucl.\ Phys.\  {\bf 56}, 340 (2006)  [astro-ph/0505266].

\bibitem{Flanagan:2006ra}
  E.~E.~Flanagan,
  Phys.\ Rev.\ D {\bf 74}, 023002 (2006)  [astro-ph/0605504].

\bibitem{Bouchami:2007en}
  J.~Bouchami and M.~B.~Paranjape,
  Phys.\ Rev.\ D {\bf 78}, 044022 (2008)  [arXiv:0710.5402 [hep-th]].

\bibitem{Maldacena:2011mk}
  J.~Maldacena,
  arXiv:1105.5632 [hep-th].

\bibitem{tHooft:2011aa}
  G.~'t Hooft,
  Found.\ Phys.\  {\bf 41}, 1829 (2011)  [arXiv:1104.4543 [gr-qc]].


\bibitem{Stelle:1976gc}
  K.~S.~Stelle,
  Phys.\ Rev.\ D {\bf 16}, 953 (1977).

\bibitem{Hinterbichler:2011tt}
  K.~Hinterbichler,
  Rev.\ Mod.\ Phys.\  {\bf 84}, 671 (2012)  [arXiv:1105.3735 [hep-th]].

\bibitem{vanDam:1970vg}
  H.~van Dam and M.~J.~G.~Veltman,
  Nucl.\ Phys.\ B {\bf 22}, 397 (1970).

\bibitem{zak}
  V.~I.~Zakharov,
  JETP Lett.\  {\bf 12}, 312 (1970)  [Pisma Zh.\ Eksp.\ Teor.\ Fiz.\  {\bf 12}, 447 (1970)].



\bibitem{deRham:2010ik}
  C.~de Rham and G.~Gabadadze,
  Phys.\ Rev.\ D {\bf 82}, 044020 (2010)  [arXiv:1007.0443 [hep-th]].


\bibitem{deRham:2010kj}
  C.~de Rham, G.~Gabadadze and A.~J.~Tolley,
  Phys.\ Rev.\ Lett.\  {\bf 106}, 231101 (2011)  [arXiv:1011.1232 [hep-th]].

\bibitem{Deser:2012qx}
  S.~Deser and A.~Waldron,
  Phys.\ Rev.\ Lett.\  {\bf 110}, no. 11, 111101 (2013)  [arXiv:1212.5835 [hep-th]].


\bibitem{Deser:2013qza}
  S.~Deser, K.~Izumi, Y.~C.~Ong and A.~Waldron,
  arXiv:1312.1115 [hep-th].




\bibitem{Alves:2009eg}
  M.~E.~S.~Alves, O.~D.~Miranda and J.~C.~N.~de Araujo,
  Phys.\ Lett.\ B {\bf 679}, 401 (2009)  [arXiv:0908.0861 [gr-qc]].

\bibitem{Newman:1961qr}
  E.~Newman and R.~Penrose,
  J.\ Math.\ Phys.\  {\bf 3}, 566 (1962).


\bibitem{Eardley:1974nw}
  D.~M.~Eardley, D.~L.~Lee and A.~P.~Lightman,
   Phys.\ Rev.\ D {\bf 8}, 3308 (1973).

\bibitem{Aasi:2013wya}
  J.~Aasi {\it et al.}  [ LIGO Scientific and  Virgo Collaborations],
  arXiv:1304.0670 [gr-qc].


\bibitem{Moon:2011gg}
  T.~Moon and Y.~S.~Myung,
   Phys.\ Rev.\ D {\bf 85}, 027501 (2012)  [arXiv:1111.2196 [gr-qc]].


\bibitem{Fleury:2013uqa}
  P.~Fleury, Helen.~Dupuy and J.~-P.~Uzan,
  Phys.\ Rev.\ Lett.\  {\bf 111}, 091302 (2013)  [arXiv:1304.7791 [astro-ph.CO]].


\bibitem{Li:2008dq}
  W.~Li, W.~Song and A.~Strominger,
  JHEP {\bf 0804}, 082 (2008)  [arXiv:0801.4566 [hep-th]].

\bibitem{Bergshoeff:2009hq}
  E.~A.~Bergshoeff, O.~Hohm and P.~K.~Townsend,
  Phys.\ Rev.\ Lett.\  {\bf 102}, 201301 (2009)  [arXiv:0901.1766 [hep-th]].

\bibitem{Bergshoeff:2013vra}
  E.~Bergshoeff, M.~Kovacevic, L.~Parra and T.~Zojer,
  PoS Corfu {\bf 2012}, 053 (2013).



\bibitem{Harmark:2007md}
  T.~Harmark, V.~Niarchos and N.~A.~Obers,
  Class.\ Quant.\ Grav.\  {\bf 24}, R1 (2007)  [hep-th/0701022].



\bibitem{Gregory:1993vy}
  R.~Gregory and R.~Laflamme,
   Phys.\ Rev.\ Lett.\  {\bf 70}, 2837 (1993)  [hep-th/9301052].

\bibitem{Myung:2014nua}
  Y.~S.~Myung and T.~Moon,
   arXiv:1401.6862 [gr-qc].

\bibitem{Myung:2013doa}
  Y.~S.~Myung,
 Phys.\  Rev.\ D {\bf 88}, 024039 (2013)  [arXiv:1306.3725 [gr-qc]].



\bibitem{Babichev:2013una}
  E.~Babichev and A.~Fabbri,
   Class.\ Quant.\ Grav.\  {\bf 30}, 152001 (2013)  [arXiv:1304.5992 [gr-qc]].

\bibitem{Brito:2013wya}
  R.~Brito, V.~Cardoso and P.~Pani,
  Phys.\  Rev.\ D {\bf 88}, 023514 (2013)  [arXiv:1304.6725 [gr-qc]].





\end{thebibliography}
\end{document}